\documentclass[12pt,preprint]{aastex} 
\usepackage{apjfonts}
\usepackage{epsfig,graphicx}
\newcommand{\varcsec}{^{\prime\prime}}

\slugcomment{Received ........; Accepted .........}

\shorttitle{Magnetic reconnection as a source of jets}

\shortauthors{Bharti et al.}

\begin{document}

\title{Magnetic reconnection as a source of jets from a penumbral intrusion into a sunspot umbra}

\author{L. Bharti$^{1}$, S. K. Solanki$^{1,2}$ and J. Hirzberger$^{1}$ }
\affil{1. Max-Planck-Institute f\"ur sonnensystemforschung, Justus-von-Liebig-Weg 3,
37077, G\"ottingen, Germany}
\affil{2. School of Space Research, Kyung Hee University, Yongin, Gyeonggi Do, 446-701, Korea}
       \email{bharti@mps.mpg.de}

\begin{abstract}
We present the results of high resolution co-temporal and co-spatial photospheric and chromospheric observations of sunspot penumbral intrusions.
The data was taken with the Swedish Solar Telescope (SST) on the Canary Islands. Time series of
Ca\,II H images show a series of transient jets extending roughly
3000 km above a penumbral intrusion into the umbra. For most of the time series
 jets were seen along the whole length of the intruding bright filament. Some of these jets develop
 a clear $\lambda$-shaped structure, with a small loop appearing at their
 footpoint and lasting for around a minute. In the framework of earlier studies, the observed transient
 $\lambda$ shape of these jets strongly suggests that they are caused by magnetic reconnection between
 a curved arcade-like or flux-rope like field in the lower part of the penumbral intrusion and the more vertical umbral magnetic
 field forming a cusp-shaped structure above the penumbral intrusion.

\end{abstract}


\section{Introduction}

High resolution observations have revealed that the chromosphere above sunspots is
very dynamic at small scales. Jet-like phenomena
observed above sunspot penumbrae (penumbral micro jets, Katsukawa et al. 2007) and above umbrae (umbral microjets, Bharti et al. 2013) in Ca\,II H are examples of such
dynamic events. The penumbral micro jets are associated with transient brightenings and
are observed at the boundaries of penumbral filaments. Their lifetime is about 1
minute and the apparent speed projected on the image plane is about 100 km/s. The length of these jets is
approximately 1000-4000 km and their width is 400 km. Jur\v c\'ak \& Katsukawa 2008 found that they are aligned with the
background magnetic field . Umbral microjets (Bharti et al. 2013) are likely also aligned with the background magnetic field.
  The typical length and width of umbral microjets is less than 1\arcsec and 0\farcs3, respectively. They last around one minute.
 Reardon et al. (2013) found that penumbral micro jets show extended emission in the wings of spectral lines , thus giving a similar signature  to Ellerman bombs (Ellerman 1917) outside sunspot.

  Other interesting dynamical phenomena in the sunspot chromosphere are  brightenings
and surge activity above  light bridges(Roy 1973, Asai et al. 2001, Berger \& Berdyugina 2003,
 Bharti et al. 2007, Louis et al. 2008, 2009, Shimizu et al. 2009, Shimizu 2011).
 In particular, Shimizu et al. (2009) and Shimizu (2011) clearly detected
 jet-like structure in light bridges. The observed bright plasma ejection was intermittent and recurrent for more than a day.
 The length of the bright ejections was reported to be 1500-3000 km, i.e. significantly less than previously reported length 10000 km of dark surges
 in H${\alpha}$ images by Roy (1973) and Asai et al. (2001). Shimizu et al. (2009) occasionally observed fan-shaped ejection
 as well as a chain of ejections from one end the of the light bridge to the other. Louis et al. (2008) reported arch-like
 structure above a light bridgeand brightness enhancement along lightbridge. Bharti (2015) finds jets above a light bridge which reach up to coronal heights and the leading edge of jets are hotter in the transition region and corona. The jets show a coordinated behaviour i.e. neighbouring jets move
  up and down together (Yang et al. 2015). Kleint \& Sainz Dalda (2013) reported on brightenings above several unusual
 filamentary structures (umbral filaments) in a sunspot. Coronal images show end points of bright coronal loops end in these unusual filaments.

 Sakai \& Smith (2008) and Magara
 (2010) proposed models for penumbral microjets that build on  magnetic reconnection taking place between
more inclined magnetic field in penumbral filaments and a more vertical background
field. A similar interpretation was proposed by Shimizu et al. (2009) for the
(bidirectional) jets emanating from a light bridge.

Chromospheric jets and surges are also found outside active regions (Shibata et al. 2007). In particular, Shibata et
al. (2007) found $\lambda$-shaped jets in the quiet chromosphere, which was interpreted to support the
notion of reconnection between an emerging magnetic bipole and a preexisting
uniform vertical field (Yokoyama \& Shibata 1995). Such a loop configuration gives indirect evidence of small-
scale reconnection in the solar atmosphere  (Singh at al. 2012a, Yan et al. 2015). Particularly, Yan et al. (2015)
found evidence for intermittent reconnection in a $\lambda$-shaped jet driven by loop advection and followed by an outflow that excites waves and jets. The authors interpret
these phenomena as causes and consequences of reconnection ,respectively.
A detailed analysis of quiet sun jets is presented by Nishizuka et al. 2011. A study of anemone jets by Morita et al. (2010)
suggests jets generated in the lower chromosphere. On the other hand a multiwavelength study of a jet show simultaneous appearance
in the lower and in the upper atmosphere (Nishizuka et al. 2008). The appearance of jets in the atmosphere depends on the size of the bipole (Shibata et al. 2007, Singh et al. 2011).

The present study is based on the high resolution G-continuum, Ca\,II H and Fe{\sc I} 6302 \AA~ observations taken with the Swedish Solar Telescope
to detect $\lambda$-shaped jets above a penumbral intrusion  into the umbra.
This letter is organized as follows: in Sec. 2 we present the observations, in Sec. 3 we describe the
$\lambda$--shaped jets and their association with the underlying photospheric umbral features, we then
present our conclusions in Sec. 4.

\section{Observations}

The observations were carried out in various
wavelength bands at the Swedish Solar Telescope (SST, Scharmer et al. 2003), La Palma, Canary
Islands on 2006 August 13.  The sunspot belonged to the active region NOAA 10904 and positioned at the heliocentric angle
$\theta= 40.15^\circ$ ($\mu = 0.76$). The sunspot was located at $x=-556\arcsec$ and $y=-254\arcsec$ on the solar disk.

The sunlight was divided into a blue and a red channels.
In the blue beam various interference filters were used to obtain
images in G-band (4305\,\AA\ ), G-continuum (4363\,\AA\ ), Ca II H (3968.5\,\AA\ ) line core and in the ling
wing (0.06\,\AA\ away from line center). The image scale in the blue beam corresponded to 0.041"/pixel.

The red beam was fed to the Solar Optical Universal Polarimeter (SOUP, see Title \&
Rosenberg 1981) filter to scan the FeI 6302.5 \AA\ line at 6 wavelength positions. The width of the filter was 75 m\AA. Full
Stokes polarimetry was performed to measure the magnetic field vector at each pixel
of the field of view. In addition, continuum (broad band) images at 6302 \AA\ were recorded.
The plate scale in the red beam corresponded to 0. 065"/pixel.  More detail on the data acquisition and
reconstruction are described in Hirzberger et al. (2009). Measurement of instrumental polarization effects of the
SST optical setup were done by inserting calibration optics, consisting of a rotating polarizer and a quarter wave plate into the beam. A code developed by Selbing (2005) was used to determine the demodulation matrix (see Hirzberger et al. (2009) for more details.)

Speckle interferometric techniques (see, Hirzberger et al. 2009) were used to
reconstructed the G-continuum, the Ca II H core and Ca II H wing time series and SOUP
polarization data. In addition, the HeLIx inversion code (Lagg et al. 2004) was used to
invert the Stokes vector data assuming a simple one-component-plus-straylight
Milne-Eddington atmosphere. The obtained magnetic parameters were transformed to the local coordinate system and
a code developed by Georgoulis (2005) was used to resolve the 180$^\circ$ ambiguity.

\begin{figure*}
\centering
\includegraphics[width=155mm,angle=0]{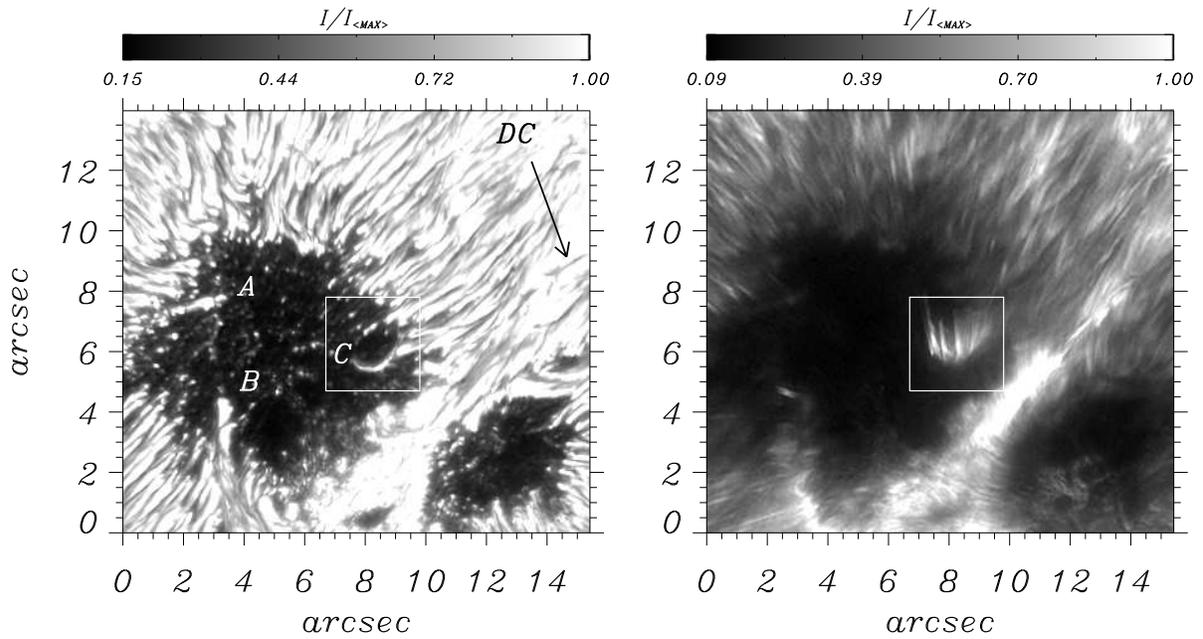}
\vspace{5mm}
\caption{Left panel: contrast-enhanced G-continuum image of the inner part of the observed sunspot at 9:11:31 UT.
 The arrow labelled 'DC' points towards disk center. Right panel: cotemporal
  and cospatial contrast-enhanced Ca II H line core image. Locations marked A, B and C in the left panel indicate
   penumbral intrusions where jet-like events are seen. White boxes outline the field-of-view
   of events discussed in more details in the text. Blow-ups of these boxes are shown in Fig. 2 and
   in the movies available as on-line material. }
\end{figure*}

\begin{figure*}

\centering
\includegraphics[width=140mm,angle=0]{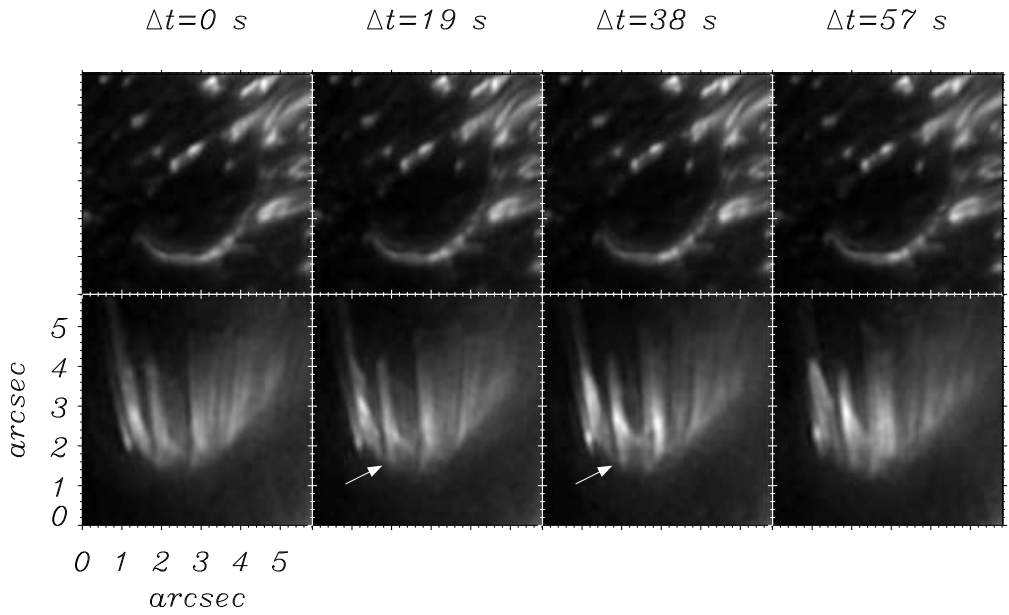}
\caption{Evolution of the jets and $\lambda$-shaped jets above penumbral intrusion C. Upper row: G-continuum images; lower row:
simultaneous and cospatial images in the Ca\,II H line core. The plotted scene
corresponds to the white squares in Fig. 1. Time $\Delta$t=0 corresponds to 9:11:12 UT (see movie-II for more details which is available as on-line material).}
\end{figure*}

\begin{figure*}
\vspace{-30mm}

\centering
\includegraphics[width=80mm,angle=0]{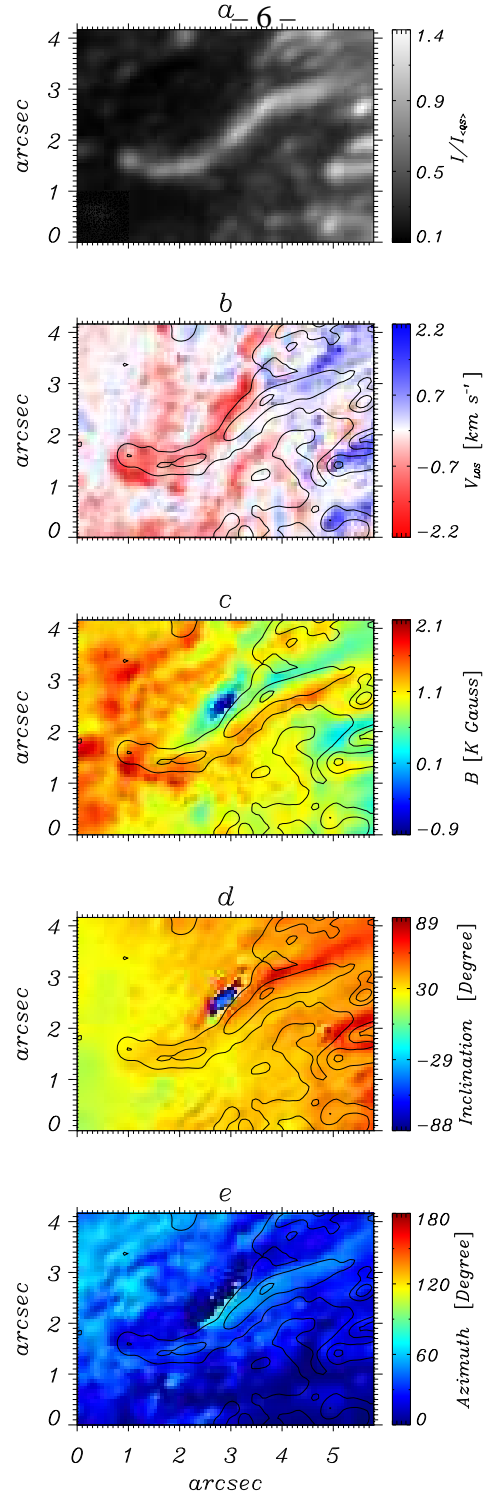}
\vspace{-10mm}
\caption{Continuum image and plasma parameters from inversions of the Stokes profiles at around 6302 \AA~. a: continuum at 6302 \AA.
b: line-of-site velocity, c: magnetic field strength, d: inclination of magnetic vector, e: azimuth.
The overplotted contours outline the umbral-penumbral boundary in the continuum images.}
\end{figure*}

\begin{figure}
\vspace{-5mm}
\hspace{28mm}
\includegraphics[width=100mm,angle=0]{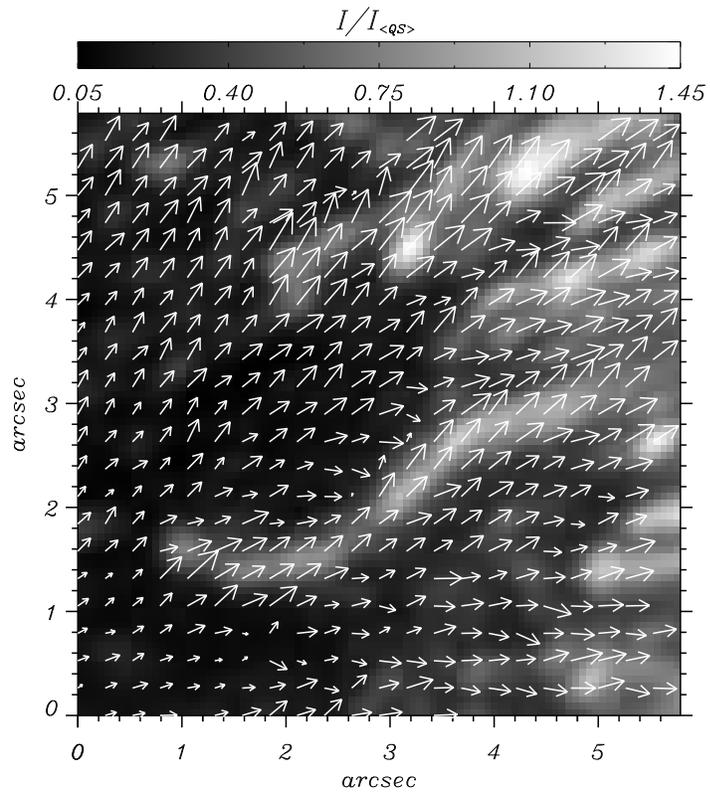}
\caption{Vectors showing the horizontal magnetic field structure overplotted on an intensity image.}
\end{figure}

\begin{figure*}
\vspace{-55mm}

\includegraphics[width=150mm,angle=0]{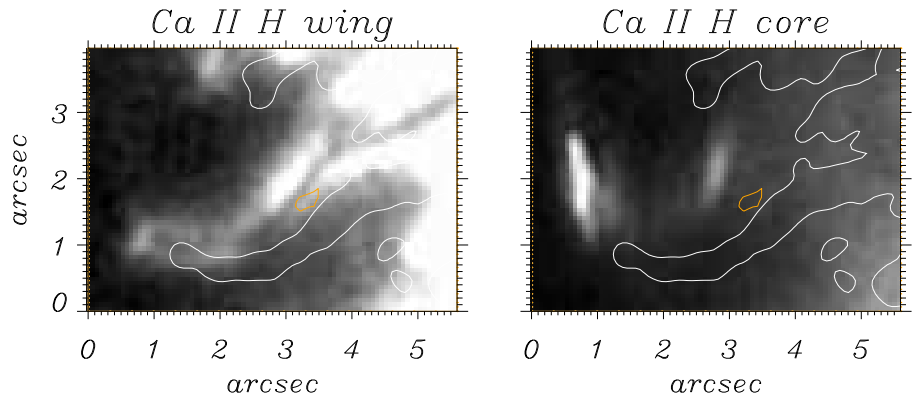}
\caption{Left: Ca\,II H wing image. An intensity threshold has been chosen to highlight the penumbral intrusion.
Right : Ca\,II H core image. The white contours correspond to a continuum (6302 \AA) threshold, the
 orange contour outlines the opposite polarity patch. All images were recorded around 08:44:55 UT.}
\end{figure*}

\begin{figure}
\vspace{-5mm}
\hspace{28mm}
\includegraphics[width=100mm,angle=0]{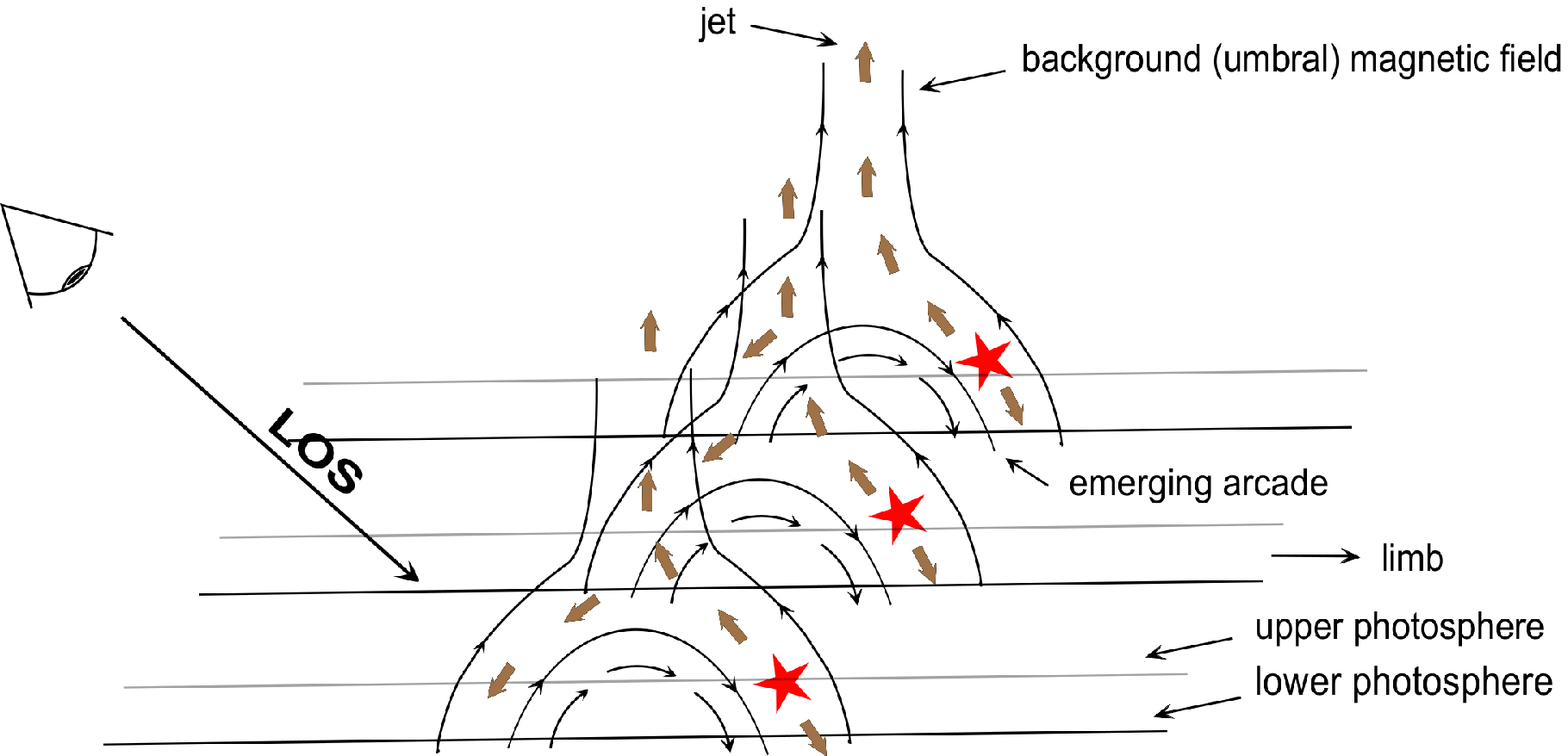}
\caption{Schematic picture of $\lambda$-shaped jets produced by magnetic reconnection between
an emerging magnetic arcade along the penumbral intrusion and the pre-existing background umbral magnetic field.
Thick arrows (brown) represent plasma accelerated by the reconnection, thin arrows indicate convective flows.
The red stars mark the location of reconnection.}
\end{figure}

\section{Analysis and result}

Contrast enhanced co-temporal and co-spatial images of the sunspot in the G-continuum
and in the Ca\,II H line core are displayed in Fig. 1. A broad light bridge(LB),
a larger and a smaller umbrae is
 visible in both the images. In the G-continuum image both the umbra show peripheral umbral dots and
several dark patches (nuclei). However, in the larger umbra central umbral dots and many penumbral
intrusions are also recognised (Bharti et al. 2013). These intrusions are also visible with lower contrast in the Ca\,II H core image.
Several jet-like transient bright structures are visible above penumbral filaments (penumbral microjets,
cf. Katsukawa et al. 2007), LB (Shimizu et al. 2009 and Shimizu 2011) and penumbral intrusions in the Ca II H core image (best seen in the mpeg movie-I which is provided as online material).

Occasionally bright jets similar to penumbral microjets are visible above penumbral intrusions A and B as can be gleaned from Movie-I.
 One can notice around 9:11 UT the jets above intrusion B have a longer life time and a larger width than jets above intrusion A.
 Most striking is penumbral intrusion C, however. There jets are both brighter and
 more extended than those from the other intrusions and occur without interruption over
 the entire observing span of about 47 min. The jets above the intrusion follow the orientation of the nearby penumbral microjets. In the following, we concentrate on the activity above penumbral intrusion C.

 In the beginning at 8:28:27 UT the jets occur only along the tip of this penumbral
 intrusion. Later on jets appear somewhat closer to the umbral-penumbral
 boundary.

 The jets' length and brightness is higher around the
 tip of the filament. Around 9:02:01 UT a bunch of jets appears near the umbral-penumbral boundary with
 comparable length and brightness as around the tip of the filament. This strong jet activity
 then rapidly migrates along the intrusion into the umbra. Apart from migration, neighbouring jets merge with each other.
 Such migration and merging of jets has been also reported above a light bridge in the transition region by Bharti (2015).

 In Fig. 2, from 9:11:12 UT on, one can
 clearly see $\lambda$-shaped jets, i.e. jets coming out of the tops of (small)loops whose footpoint lie in or next to the penumbral
 intrusion. Later, after 9:12:28 UT, again only elongated jets are visible in subsequent frames, although
 a weak $\lambda$-like structure is still hinted at.

 In the G-continuum images the penumbral intrusion shows a dynamical behaviour reminiscent of twisting motions and
 migrations of bright grains (see mpeg Movie-II). A clear association of
 bright penumbral grain migration towards the umbra and the jets is visible around the tip of
 the penumbral intrusion from 8:42:23 UT to 8:45:52 UT. Similarly, from 9:11:12 UT
 onwards (i.e. starting with the appearance of $\lambda$-shaped jets), migration of bright penumbral
 grains towards the umbra in G-continuum images close to the footpoints of $\lambda$-shaped
 jets can be recognised.

      We applied an intensity  threshold (c.f. Bharti et al. 2013 for more details) to
      determine various parameters of the jets. The projected lengths
      of jets are found to be 1700-3000 km. The $\lambda$-shaped jets belong to the shorter ones
      with lengths of 1700-1900 km. The separation between the footpoints of $\lambda$-shaped jet
      is 450-600 km and the loop at the bottom of the $\lambda$ structure reaches a height of around 600 km
      above the loop base. These jets have shorter length than penumbral microjets (Katsukawa et al. 2007) but longer length  than umbral microjets (Bharti et al. 2013). The lifetime of these jets (2-3 min) is comparable with penumbral
      microjets (Katsukawa et al. 2007) and umbral microjets (Bharti et al. 2013) as well as with jets reported in umbra by
      Yurchyschyn et al. (2014).

  Panel 'a' of Fig. 3 display the continuum intensity (broad-band at 6302 \AA) at
  8:44:31 UT. The penumbral intrusion appears similar to that in the G-continuum (see Movie-II),
  but, due to the somewhat lower spatial resolution and contrast (caused by the longer wavelength), the fine structures are less clear. The
  line-of-sight velocity is depicted in panel 'b'. The part of the intrusion close to the penumbra-umbra boundary
  shows upflows, similar to the ones typical of the inner penumbra while the part away from the penumbra-umbra boundary shows upflows in the section
  towards disk center and downflows away from disk center. From $x=1.5\varcsec$ towards the tip of the intrusion
  only downflows are visible. The magnetic field strength is plotted in panel 'c'. It is
  weaker in the intrusion compared to the surroundings, but always above 1 kG (in the mid-photosphere), again similar to filaments embedded
  in the penumbra. Panel 'd' suggests that the field in the intrusion is more transverse, where the direction perpendicular to the solar surface refers to zero degrees. Interestingly, there is
  an opposite polarity patch, around $x=2.7\varcsec$ and $y=1.8\varcsec$, at the edge of the
  intrusion, close to the penumbra-umbra boundary. The azimuth (panel 'e') also shows about 90$^{\circ}$
  rotation compared to the rest of the field surrounding the opposite polarity
  patch. The positive y axis is taken to correspond to a field azimuth of zero. The arrows overplotted on the intensity image in Fig. 4 show the horizontal field. Note that there may be smaller amounts of opposite polarity flux also along other parts of the filament hidden by the dominant umbral field.
  Although we observe patches of opposite polarity, we are aware of the fact that with 6 wavelength points and highly redshifted profiles, this interpretation
  is not straightforward. Since there are observational hints (Esteban Pozuelo et al. 2015) and well-established results from simulations (Rempel 2012) for the existence of opposite polarity field at the edges of penumbral filaments as well as in light bridges (Bharti et al. 2007, Lagg et al. 2014, Louis et al. 2014) the observed opposite polarity patches might be real. The presence of jets beyond the opposite polarity
  patch support this notion.

The left panel of Fig. 4 displays a Ca\,II\,{\it H} wing image of the penumbral intrusion at 08:44:55
UT. The umbral-penumbral boundary, and the penumbral intrusion as visible in the continuum image (6302 \AA),
are marked by white contours. The orange contour
outlines the opposite polarity patch. In the Ca\,II H wing image the penumbral intrusion shows a
filamentary structure with a central dark lane along its whole length. In continuum radiation, formed somewhat deeper  than the line wing, the intrusion is much narrower and only the bright part adjacent to the central dark lane toward disk center is clearly visible. The other
part, on the limbward side of the central dark lane, is hardly seen. Note that the sunspot is located away from the disc
center at $\theta= 40.15^\circ$, and the intrusion corresponds to an elevated structure (Lites et al.
2004) due to the raised optical depth unity level(Cheung et al. 2010). It is unclear if this 3-D structure of the iso-$\tau$
surface is responsible for the difference in visibility and structure of the intrusion at different wavelengths.
The velocities may also give an indication why the intrusion looks so different in the lower photosphere
than in the middle and upper photosphere. The weak upflow in the disc center side (i.e. in the continuum intrusion) and the strong downflows
in the part close to the limb (visible only in the Ca wing) can be described as in Fig. 5. We propose that these flows have different physical causes:
a) convection (narrow arrows in Fig. 5), with upflows on the disc center side of the arcade and downflows on the other side. These convective flows heat
the gas in the lower photosphere, but only at the location of the upflows, producing a narrow intrusion in lower photosphere.
b) in the upper photosphere and the chromosphere, the intrusion is heated more by the reconnection, which accelerates the gas in both directions (thick
arrows in Fig. 5) and thus brightens both parts of the intrusion located at the footpoints of the emerging arcade. The fact that the reconnection takes place closer to the limbward footpoint of the arcade may explain why the intrusion is brighter in Ca wing on the limbward side at many places. Due to increasing
gas density with depth, the reconnection does not affect the lower photosphere. The width of the penumbral intrusion in the 6302 \AA~ continuum images is 400-550 km, which is comparable to, but somewhat narrower than
the footpoint separation of the $\lambda$-shaped jets in the Ca\,II H line core images (see Fig. 2). The width of the intrusion, as
seen in the wing of Ca\,II H, is sufficient, however, to easily host both legs of the $\lambda$-shaped jets.
The right panel of Fig. 4 depicts a Ca\,II  H line core image with
overplotted contours of the continuum (6302 \AA) and the opposite polarity patch as in
the left panel. At $x=2.7\varcsec$ and $y=1.8\varcsec$, a jet can be seen which seems to emerge directly from the
opposite polarity. Close inspection of Movie-I shows that the location of this jet activity
 corresponds to $x=8.5\varcsec$  and  $y=6\varcsec$ in Ca\,II H that peaked at 08:44:55 UT. It is evident from Fig. 3 and 4 that there is also a downflow and opposite polarity patch
  surrounding $x=2.7\varcsec$ and $y=1.8\varcsec$. This location
  corresponds to the footpoints of the jets seen in Ca\,II H line core. Louis et al. (2014) found small scale jets above
  light bridges in Ca\,II H line core images from Hinode (SOT) observations which are associated with localized patches of opposite polarity
  in the photosphere. The jets are triangular-shaped and exhibit a spike-like structure. The width of the filter used in the present study is
  narrower (1.1 \AA) than the one used in Hinode (SOT/BFI). In addition, the spatial resolution of the present observations is nearly a factor of two higher, which might enable us to unravel the sub-structure ($\lambda$-shape)
  of the jets.

\section{Discussion and conclusion}

We have for the first time clearly detected $\lambda$-shaped jets in a sunspot. These
lie above a penumbral intrusion into the umbra. With a length of 1800 km they are at the small
end of the quiet sun $\lambda$-shaped jets (lengths of 2000-5000 km) reported by Shibata et al.
(2007), based on Hinode/SOT Ca\,II H observations. This shape of a jet is
considered to be a signal of reconnection between the loop connecting an emerging
magnetic bipole and preexisting more vertical field (Yokoyama and Shibata 1995, Shibata et al. 1992, 2007).
  In our case a whole arcade of
emerging flux is required, as we see jets all along the intrusion, although not all of
them display a $\lambda$-type structure possibly due to overlap between jets or
insufficient spatial resolution. As can be deduced from the movie-II,
 the group of jets to which the $\lambda$-shaped jets belong moves rapidly along
the penumbral intrusion, starting at the umbral-penumbral boundary and
ending at the tip of the intrusion. This implies an emerging arcade and suggests that the arcade first
started  to emerge near the umbra-penumbra boundary and later towards
the tip of the intrusion. This notion is supported by the fact that we also see
a migration of bright penumbral grains (in the G-continuum) toward the tip of the intrusion
and close to the $\lambda$-shaped jets. Instead of an arcade, a rising and heavily twisted flux
rope (and possibly rotating) could also be the cause the jets, or a strong elongated
convective upwelling, carrying opposite polarity magnetic flux to its edges. Singh et al. (2012)
found systematic motions of $\lambda$-shaped jets outside a sunspot in Ca II H observations from Hinode in terms of migration from
one end of the footpoint to the other end of an arcade that, finally, leads to merging of jets similar to the merging of jets in the peumbral
intrusion presented here. Such a morphology is illustrative of the emergence of twisted flux rope (see Fig. 4 of Singh et al. 2014).
The presence of the $\lambda$-shaped loops, however, requires a relatively ordered loop-like field, as
illustrated in Fig. 5. This scenario is reproduced by a rising arcade of loops connected with
convective upwelling. The convective upflow heats the gas in the lower photosphere thus loops connected with convective upwelling brightens the disk side
penumbral intrusion in the continuum. The appearance of the bright $\lambda$-shaped loop in the upper photosphere is due to
the heating caused by the reconnection, which accelerates the gas in both directions, filling both the loop with hot and bright gas as well as the jet above it.

   Magara (2010) performed a MHD simulation to explain penumbral microjets.
  In his model a strongly twisted flux rope (penumbral filament) is placed within the more vertical
  background field (umbral field). Magnetic reconnection takes place on the
  side of the flux rope with field lines having opposite polarity to the
  background field. Bharti et al. (2010) found reverse polarities at edges of larger UDs, in the
 simulations of Sch\"ussler \& V\"ogler (2006), caused by strong convective downflows.
 The presence of opposite polarity along the lateral edges
  of penumbral filaments has been reported recently by Ruiz Cobo \& Asensio
  Remos (2013), Scharmer et al. (2013), Tiwari et al. (2013) and Esteban Pozuelo et al. (2015). These opposite polarities
  are thought to be caused by convective downflows (Joshi et al. 2011; Scharmer et al. 2011). The observed up and downflows pattern, their correlation
  with lower photospheric brightness (bright upflows, dark downflows) opposite polarity and
  (possibly to a lesser extent) apparent twisting motions
  in the penumbral intrusion similar to penumbral filaments (Ichimoto at al 2007, Zakharov et al. 2008, Bharti et al. 2010, 2012)
  suggests that the intrusion is of convective nature similar to penumbral filaments, umbral dots and light bridges (Cheung et al. (2010).
  This supports the reconnection scenario of
  Magara (2010), although the full structure of the magnetic field may be different from what he
  proposed to produce a jet-like structure (see Tiwari et al. 2013). The approximate alignment of
  penumbral microjets with the background field found by Jur\v c\'ak, \& Katsukawa (2008)
  was explained by Nakamura et al. (2012) as caused by reconnection between a
  weaker more horizontal and a stronger background field. Shimizu et al. (2009)
  proposed a model to interpret plasma ejections above a LB.
  In their model the current-carrying and highly twisted LB field is trapped below a
  cusp-shaped magnetic structure formed by the background field. This geometry is then proposed to
  lead to reconnection on the side of the LB at which the field is of opposite polarity to the
  umbral field. In a laboratory experiment Nishizuka et al. (2012) succeeded in producing jets
  that are qualitatively similar to penumbral jets as a result of magnetic reconnection
  in a roughly similar magnetic configuration. The general magnetic configuration presented by Magara (2010)
 and Shimizu et al. (2009) is in general agreement with our findings for the penumbral intrusion. However,
 different sources of the magnetic flux forming the  arcade are possible. It may be an emerging arcade of loops along the
intrusion, or an emerging or twisting flux rope. Finally, fields twisted and
dragged down by magnetoconvection are also candidates. Since
the $\lambda$ jets are seen in chromospheric radiation and the loop at the base of the jet
is around 600 km high, we expect that the reconnection is taking place in the
upper photosphere.

\acknowledgments

We thank an anonymous referee for helpful comments that improved this paper.
The Swedish 1-m Solar telescope is operated on the island of La Palma by the
Institute for Solar Physics of the Royal Swedish Academy of sciences in the
Spanish Observatorio del Roque de los Muchachos of the Instituto de
Astrof\'isica de Canarias. L.B. is grateful to the Inter University
Centre for Astronomy and Astrophysics (IUCAA) Reference
Center at the Department of Physics, Mohanlal Sukhadia University,
Udaipur, India, for providing computational facilities. This work has been partly
supported by the BKZI plus program through the NRF funded by the Korean Ministry of
Education.

\end{document}